\documentstyle[12pt]{article}

\def\overlay#1#2{\ifmmode%
\setbox0=\hbox{$#1$}%
\setbox1=\hbox to\wd0{\hss$#2$\hss}\else%
\setbox0=\hbox{#1}%
\setbox1=\hbox to\wd0{\hss#2\hss}\fi%
 #1\hskip-\wd0\box1 }

\catcode`@=11
\newcount\@tempcntc
\def\@citex[#1]#2{\if@filesw\immediate\write\@auxout{\string\citation{#2}}\fi
  \@tempcnta\z@\@tempcntb\m@ne\def\@citea{}\@cite{\@for\@citeb:=#2\do
    {\@ifundefined
       {b@\@citeb}{\@citeo\@tempcntb\m@ne\@citea\def\@citea{,}{\bf ?}
\@warning
       {Citation `\@citeb' on page \thepage \space undefined}}%
    {\setbox\z@\hbox{\global\@tempcntc0\csname b@\@citeb\endcsname\relax}%
     \ifnum\@tempcntc=\z@ \@citeo\@tempcntb\m@ne
       \@citea\def\@citea{,}\hbox{\csname b@\@citeb\endcsname}%
     \else
      \advance\@tempcntb\@ne
      \ifnum\@tempcntb=\@tempcntc
      \else\advance\@tempcntb\m@ne\@citeo
      \@tempcnta\@tempcntc\@tempcntb\@tempcntc\fi\fi}}\@citeo}{#1}}
\def\@citeo{\ifnum\@tempcnta>\@tempcntb\else\@citea\def\@citea{,}%
  \ifnum\@tempcnta=\@tempcntb\the\@tempcnta\else
   {\advance\@tempcnta\@ne\ifnum\@tempcnta=\@tempcntb \else
\def\@citea{--}\fi
    \advance\@tempcnta\m@ne\the\@tempcnta\@citea\the\@tempcntb}\fi\fi}
\catcode`@=12

\begin{document}
\input epsf

\hfill
{\vbox{
\hbox{CPP-95-8}
\hbox{DOE-ER-40757-066}
\hbox{May 1995}}}

\begin{center}
{\large \bf Recent Progress on Perturbative QCD Fragmentation Functions}

\vspace{0.1in}

Kingman Cheung\footnote{Invited talk at PASCOS/HOPKINS 1995 Symposium,
Johns Hopkins University, Baltimore, Maryland, March 22--25, 1995.}

{\it Center for Particle Physics, University of Texas at Austin,
Austin TX 78712 U.S.A.}

{\small email: cheung@utpapa.ph.utexas.edu}
\end{center}

\begin{abstract}
The recent development of perturbative QCD (PQCD) fragmentation functions
has strong impact on quarkonium production.
I shall summarize $B_c$ meson production based on these PQCD fragmentation
functions, as well as, the highlights of some recent activities on applying
these PQCD fragmentation functions to explain anomalous $J/\psi$ and $\psi'$
production at the Tevatron.   Finally, I discuss
a fragmentation model based on the PQCD fragmentation functions
for heavy quarks fragmenting into heavy-light mesons.
\end{abstract}

\section{Introduction}

One of the biggest ever discrepancies between theoretical predictions
and experimental
data was $J/\psi$ and $\psi'$ production observed by CDF \cite{cdf}
at the Tevatron.   With the recent development of perturbative QCD (PQCD)
fragmentation functions the experimental data can be accommodated within
reasonable uncertainties.  In the following I shall briefly describe the
essence of PQCD fragmentation functions \cite{gfrag,cfrag,bcfrag}, then
summarize $B_c$ meson production based on these PQCD fragmentation
functions \cite{prl,induce,total} and the highlights of some recent
activities on applying the PQCD fragmentation functions to explain
anomalous $J/\psi$ and $\psi'$ production
\cite{CG,BDFM,roy-1,cho,cho-spin,close,roy-2,sean}.
Finally, I shall discuss a fragmentation model \cite{spin,talks,pqcd} based
on these PQCD fragmentation functions for the fragmentation of heavy
quarks into heavy-light mesons, {\it e.g.}, $c\to D,D^*$ and $\bar
b\to B,B^*$.  This model is more attractive than previous
fragmentation models since it is based on PQCD and the PQCD
fragmentation functions have the correct heavy quark behavior.

\begin{figure}[t]
\centering
\leavevmode
\epsfysize=150pt
\epsfbox{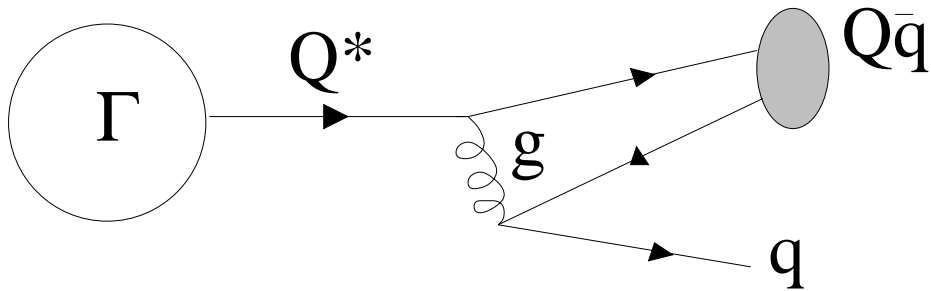}
\caption[]{
\label{fig1} \small
Schematical diagram for a heavy quark $Q$ fragmenting into
$(Q\bar q)$ and $q$.
}
\end{figure}

In general, fragmentation of quarks and gluons lies in the nonperturbative
regime so that the fragmentation functions cannot be calculated
{}from first principle.  But there is a particular
class of fragmentation functions,
namely those for heavy quarks or gluons fragmenting into heavy-heavy
bound-states, that is calculable in PQCD.  Heavy-heavy bound-states refer
to heavy quarkonia, $(c\bar c)$, $(b\bar b)$, and $(\bar b c)$ mesons.
To visualize let us consider
the hadronization of a heavy quark $Q$ into a meson $Q\bar
q$, which is schematically shown in Fig.~\ref{fig1}.  It is the lowest
order 1-gluon exchange diagram that describes the hadronization process.
As shown in Fig.~\ref{fig1}, the central part is the creation of the
quark-pair $q\bar q$ out of the vacuum, followed by the binding of $\bar q$
to $Q$ to form the meson $Q\bar q$.  Therefore, the natural scale
of this process is of order $m_q$, specifically  we choose it
to be $2m_q$.  Figure \ref{fig1} could be used to picture the
fragmentation of a heavy quark into a heavy-light meson when $Q=b,c$
and $q=u,d$.  The natural scale is then equal to $2m_{u,d}$, which
is close to $\Lambda_{\rm QCD}$, so that we do not expect
the process to be calculable by PQCD.  The
nonperturbative physics involved demands a model calculation.

On the other hand,  figure \ref{fig1} can also
be used to describe the production
of $(b\bar b)$, $(c\bar c)$, and $(\bar b c)$ mesons when $Q=b,c$ and $q=b,c$.
But the main difference is that the natural scale of the process is now of
order $2m_{b,c}$, which is much larger than $\Lambda_{\rm QCD}$.  In other
words, the fragmentation process can be reliably calculated as an
expansion in $\alpha_s(2m_{b,c})$.  While the diagram in Fig.~\ref{fig1}
represents the lowest order $\alpha_s^2$ term; higher order corrections
can be systematically calculated.
The arguments for the calculability of gluon fragmentation into
charmonium and bottomium within PQCD are essentially the same.

Even though we can calculate the fragmentation process by PQCD, there
are bound-state effects that have to be taken care of.  The nonperturbative
bound-state effects can be parameterized by, {\it e.g.}, wavefunctions or
derivatives of wavefunctions of the bound-states at the origin.
For the case of S-wave mesons there is only one nonperturbative parameter --
$R(0)$, the wavefunction at the origin; while for P-wave mesons there
are two nonperturbative  parameters, which correspond to the color-singlet
and color-octet mechanisms.  Fragmentation into different spin-orbital
states can be obtained by using the appropriate spin projections.

The $b,c$ quark fragmentation functions can be calculated by this expression
\begin{equation}
\label{basic}
D_{Q \to Q\bar q}(z)
\;=\; \frac{1}{16\pi^2} \;\int ds
\; \theta \left( s - {M^2 \over z} - {m_q^2 \over 1-z} \right)
\lim_{p_{Q_0}/m_Q \rightarrow \infty} {|{\cal M}|^2 \over |{\cal M}_0|^2} \;
\end{equation}
where $Q,q=b,c$,
$M=m_Q+m_q$ is the mass of the meson, ${\cal M}$ is the amplitude for
producing a $Q\bar q$ and $q$ from an off-shell $Q^*$ with
virtuality $s=p_Q^2$ (Fig.~\ref{fig1}),
$p_Q$ is the 4-momentum of the heavy quark $Q$,
and ${\cal M}_0$ is the amplitude for producing the heavy quark  $Q$
with the same 3-momentum $\vec p_Q$.
Here I only present the results for $\bar b \to (\bar bc)$ in the S-wave
states \cite{bcfrag}:
\begin{eqnarray}
\label{dz1}
D_{\bar b\rightarrow \bar bc(^1S_0)}(z,\mu_0) & = &
N \, \frac{rz(1-z)^2}{(1-(1-r)z)^6}
\left[ 6 - 18(1-2r)z + (21 -74r+68r^2) z^2 \right. \nonumber \\
 && \left. -2(1-r)(6-19r+18r^2)z^3  + 3(1-r)^2(1-2r+2r^2)z^4 \right]
\end{eqnarray}
for the ground state denoted by $B_c$, and
\begin{eqnarray}
\label{dz2}
D_{\bar b\rightarrow \bar bc(^3S_1)}(z,\mu_0) & = &
3N\,  \frac{rz(1-z)^2}{(1-(1-r)z)^6}
\left[ 2 - 2(3-2r)z + 3(3 - 2r+ 4r^2) z^2  \right . \nonumber \\
&& \left.  -2(1-r)(4-r +2r^2)z^3  + (1-r)^2(3-2r+2r^2)z^4 \right]
\end{eqnarray}
for the first excited state denoted by $B_c^*$, where
$N=2\alpha_s(2m_c)^2 |R(0)|^2/(81 \pi m_c^3)$ and $r=m_c/(m_b+m_c)$.
The results for longitudinally and transversely polarized $B_c^*$ mesons
can be found in Ref.~\cite{spin}, and the results for P-wave can be found
in Refs.~\cite{pbcfrag,chen}.  The fragmentation functions for charm quark
into charmonium can be obtained by putting $r=1/2$ in the above expressions,
and  can be found in Ref. \cite{cfrag,pbcfrag,falk}.
Gluon fragmentation functions into charmonium and bottomium can be calculated
in a similar fashion, and the results are  in
Ref.~\cite{gfrag} for S-wave, Ref.~\cite{gpfrag} for P-wave, and
Ref.~\cite{cho-D} for $^1D_2$ (see also Refs.~\cite{ma} for a derivation
{}from a field theoretical definition).
Likewise,  the photon and lepton fragmentation functions into charmonium
were calculated in Ref.~\cite{sean-fg}.

A couple of remarks about these PQCD fragmentation functions are in order.
(i) The scale of the PQCD fragmentation functions calculated should be
of order of the heavy quark mass $m_Q$.  We choose it to be $m_Q + 2m_q$,
which is the minimum virtuality of the fragmenting quark.
In the case of gluon fragmentation functions, the
scale is set at $3m_Q$.
These fragmentation functions obey the usual Altarelli-Parisi evolution
equations such that fragmentation functions at higher
scale can be obtained by evolving the Altarelli-Parisi equations.
(ii) The inputs to these PQCD fragmentation functions are simply the
masses of charm and bottom quarks, and the wavefunctions at the origin
(also derivatives of the wavefunctions and color-octet
wavefunctions for higher orbital states).  All of these quantities
can be reliably obtained from potential models or from lattice simulations.
Since the inputs can be reliably obtained from other sources, these PQCD
fragmentation functions have high predictive power.

In Sec. II the fragmentation approach for calculating the $B_c$ meson
 production is described, and the same approach has been used to calculate
the fragmentation contribution to the $\psi$ and
$\psi'$ production, as will be
summarized in Sec. III.  Sec. IV describes a fragmentation model.

\section {$B_c$ Meson Production}

The fragmentation approach has been used in calculating the fragmentation
contribution to the production of $\bar bc$, charmonium, and bottomium.
This approach is based on factorization, in which
the production process is separated into the production of high energy
partons (quarks and gluons) and the fragmentation of these partons into
the meson.
In this section, I summarize the results from a series of studies
\cite{prl,induce,total} on the production rates of the S-wave and P-wave
$(\bar bc)$ mesons, as well as, the inclusive production of $B_c$.
$(\bar bc)$ mesons belong to another
heavy-quark bound state family, which is made up of a $\bar b$ antiquark
and a $c$ quark.  The spectroscopy for the spin-orbital states is similar
to that of charmonium and bottomium, and $(\bar bc)$ can be obtained by
interpolating between charmonium and bottomium.
According to potential models \cite{quigg}, the 1S, 1P, 1D, 2S,
and possibly the whole set of 2P states lie below the BD threshold.
A very peculiar feature of the excited $(\bar b c)$ states is that the
annihilation channel is suppressed relative to the electromagnetic
or hadronic transitions into lower-lying states.  Therefore, when an
excited state is produced it will cascade into the ground state with
emission of photons and pions.   Hence, all the 1S, 1P, 1D, 2S, and probably
2P states contribute
to the inclusive production of the ground state $B_c$ meson.
The D-wave fragmentation functions are not available yet but they are
expected to contribute only a  very small fraction.

The differential cross section for
producing a $(\bar bc)$ meson in a spin-orbital state $H$ is given by
\begin{eqnarray}
\frac{d \sigma}{d p_T}(p\bar p \to  H(p_T) X) & = & \sum_{ij}
\int dx_1 dx_2 dz f_{i/p}(x_1,\mu) f_{j/\bar p}(x_2,\mu)
\left [
\frac{d \hat \sigma}{dp_T} (ij \to \bar b(p_T/z)X,\,\mu) \nonumber \right. \\
&& \left. \times
D_{\bar b \to H} (z,\mu)
+ \frac{d\hat \sigma}{dp_T} (ij\to g(p_T/z)X,\mu)\; D_{g\to H}
(z,\mu) \right ] \; ,
\label{*}
\end{eqnarray}
where $f_{i/p}(x)$'s are the parton distribution functions, $d\hat \sigma$'s
are the subprocess cross sections, and $D_{i\to H}(z,\mu)$'s represent
the parton fragmentation functions at the scale $\mu$.  The factorization
scale $\mu$ is chosen in the order of the $p_T$ of the parton, so as to
avoid large logarithms in $d\hat \sigma$, and the resulting large
logarithms of order $\mu/m_b$ in $D_{i\to H}(z,\mu)$ can be summed by
the Altarelli-Parisi equations.
The gluon fragmentation functions at the initial scale
are $\alpha_s$ suppressed relative to the $\bar b$ fragmentation functions,
so we simply take the initial gluon fragmentation functions to be zero.
This is justified since the majority of the gluon fragmentation
comes from the Altarelli-Parisi evolution, and we called it the induced
gluon fragmentation functions \cite{induce}.
The $\bar b$ and gluon fragmentation functions into a $(\bar bc)$ state $H$
satisfy the following evolution equations
\begin{equation}
\label{Db}
\mu \frac{\partial}{\partial \mu} D_{\bar b\to H}(z,\mu) =
\int_z^1 \frac{dy}{y}
P_{\bar b\to \bar b}(z/y,\mu)\; D_{\bar b \to H}(y,\mu) +
\int_z^1 \frac{dy}{y} P_{\bar b\to g}(z/y,\mu)\; D_{g \to H}(y,\mu) \,,
\end{equation}
\begin{equation}
\label{Dg}
\mu \frac{\partial}{\partial \mu} D_{g\to H}(z,\mu) = \int_z^1 \frac{dy}{y}
P_{g \to \bar b}(z/y,\mu)\; D_{\bar b \to H}(y,\mu) +
\int_z^1 \frac{dy}{y} P_{g \to g}(z/y,\mu)\; D_{g \to H}(y,\mu) \,,
\end{equation}
where $P_{i\to j}$ are the usual
Altarelli-Parisi splitting functions.
The initial conditions to the above equations are simply the initial $\bar b$
fragmentation functions
and the initial gluon fragmentation functions, which are set to zero.
We can also examine the relative importance of these fragmentation functions.
The initial $D_{\bar b\to H}(z,\mu_0)$ is of order $\alpha_s^2$, even when
it is evolved to a higher scale it is still of order $\alpha_s^2$.
In contrast, the initial $D_{g\to H}(z,\mu_0)$ is of order $\alpha_s^3$,
but at a higher scale $\mu$ it is of order $\alpha_s^3 \ln (\mu/\mu_0)$
via the Altarelli-Parisi evolution.  Therefore, at a sufficiently large scale
the induced $D_{g\to H}(z,\mu)$ is as important as the $\bar b$ quark
fragmentation.

\begin{figure}[t]
\centering
\leavevmode
\epsfysize=200pt
\epsfbox{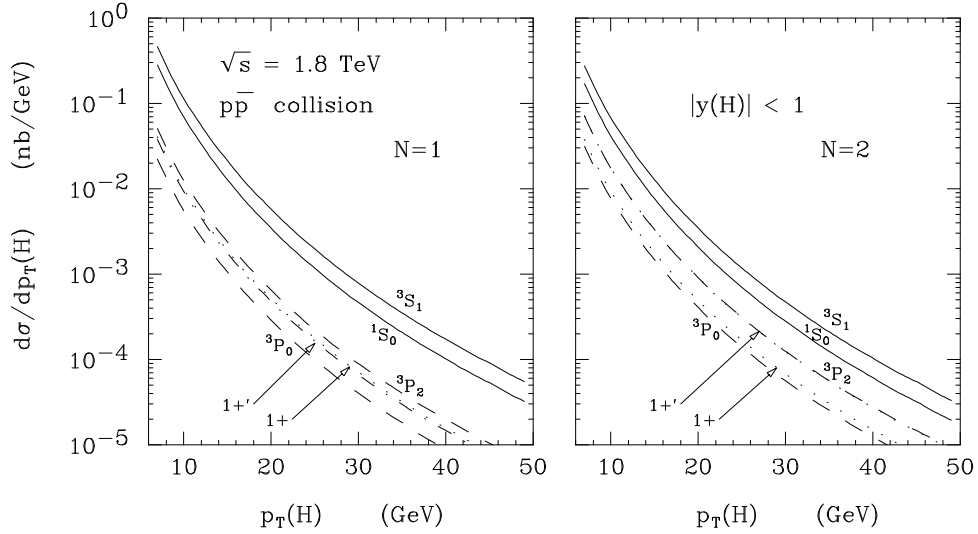}
\caption[]{
\label{fig2} \small
Differential cross sections versus the $p_T$ of the $(\bar bc)$ mesons in
various spin-orbital states with $N=1$ and $N=2$, respectively.  The
acceptance cuts are $p_T(H)>6$ GeV and $|y(H)|<1$.
}
\end{figure}

The resulting $p_T$ spectra for the S-wave and P-wave states are shown in
Fig.~\ref{fig2}.  We can now predict the inclusive production rate of the
ground state $B_c$ meson.  We add up the cross sections from all
S-wave and P-wave production, and the inclusive cross sections of $B_c$ as
a function of $p_T^{\rm min}(B_c)$ is shown in Table \ref{table1} at the
Tevatron.
Variations with factorization scale between $\mu_R/2$ to $2 \mu_R$, where
$\mu_R= \sqrt{p_T^2({\rm parton})+m_b^2}$, are also illustrated.
The variation is at worst a factor of two, and substantially reduced
at $p_T^{\rm min}>10$ GeV.
With a production cross section of about 5 nb and 100 pb$^{-1}$ integrated
luminosity at the Tevatron there are about
$5\times 10^5 \; B_c^+$ mesons.  The detection mode for the $B_c$ meson
will be $B_c \to J/\psi + X$, where $X$ can be a $\pi^+$, $\rho^+$,
or $\ell^+ \nu_l$, and $J/\psi$ can be detected easily through its
leptonic decay modes.
When $X$ is $e^+\nu_e$ or $\mu^+\nu_\mu$, it will be a striking
signature of three-charged leptons coming  from a common
secondary vertex.  The combined branching ratio of
$B_c\to J/\psi \ell^+\nu_\ell \to \ell'^+ \ell'^- \ell^+ \nu_\ell\;
(\ell,\ell'=e,\mu)$ is about 0.2\%.  This implies that there will be of order
$10^3$ such distinct events for 100 pb$^{-1}$ luminosity at the Tevatron.
However, this mode does not afford  the full reconstruction of the $B_c$.
If $X$ is some hadronic states, {\it e.g.}, pions, the events can be
fully reconstructed and the $B_c$ meson mass can be measured.
The process $B_c\to J/\psi+ \pi^+ \to \ell^+\ell^- \pi^+$ is  likely to be
the discovery mode for $B_c$. Its combined branching ratio is
about 0.03\%, which implies about 300 such distinct
events  at the Tevatron with a luminosity of  100 pb$^{-1}$.
The production rates of $B_c$ at the LHC will be of order $10^9$ with
100 fb$^{-1}$ luminosities, promising a very exciting experimental
program on $B_c$ mesons at the LHC.

\begin{table}[b]
\caption[]{\small The  inclusive production cross sections for
the $B_c$ meson at the Tevatron including the contributions from all the
S-wave and P-wave states below the $BD$ threshold as a function
of $p_{T}^{\rm min}(B_c)$.
The acceptance cuts are $p_T(B_c)>6$ GeV and $|y(B_c)|<1$.
\label{table1}
}
\bigskip
\centering
\begin{tabular}{c@{\extracolsep{0.5in}}ccc}
\hline
\hline
 $p_{T}^{\rm min}$ (GeV)  &  \multicolumn{3}{c}{$\sigma$ (nb)} \\
\hline
     &\underline{$\mu = \frac{1}{2}\mu_R$} & \underline{$ \mu=\mu_R$}
 & \underline{$\mu=2 \mu_R$} \\
 6         &  2.81  &  5.43  &  6.93 \\
10         &  0.87  &  1.16  &  1.22 \\
15         &  0.26  &  0.29  &  0.26 \\
20         &  0.098 &  0.097 &  0.083\\
\hline
\end{tabular}
\end{table}

There also exist complete ${\cal O}(\alpha_s^4)$ calculations on the
production of S-wave $B_c$ mesons \cite{bc-pqcd,ph-bc}.
There is a controversy whether the
fragmentation contribution will dominate the production of $B_c$ mesons
at the large $p_T$ regions over the non-fragmentation (recombination)
contribution.  The controversy arises
because the set of  fragmentation diagrams is a gauge-invariant subset
of the whole set of Feynman diagrams at the order $\alpha_s^4$.
So there is a competition between the fragmentation diagrams and the
recombination diagrams.  Nevertheless, the
bottom line is that the fragmentation approach
identifies the correct scale for the fragmentation diagrams, and
should give a lower bound on the production cross section of $B_c$ mesons.

\section{$J/\psi$ and $\psi'$ Production}

In this section, I highlight some recent activities on $J/\psi$ and
$\psi'$ production at the Tevatron.  Before the recent CDF data \cite{cdf}, the
dominant mechanism of $J/\psi$ production at high $p_T$ region
was believed to be
the fusion mechanism, $gg\to \chi_{cJ} g$, followed by the radiative decay
of $\chi_{cJ} \to J/\psi+\gamma$, while for $\psi'$ production the dominant
mechanism is $gg\to \psi' g$ because of the absence of $\chi_{cJ}(2P)$
states.
But the CDF measurements on $J/\psi$ and $\psi'$ exposed large
discrepancies between theoretical predictions and experimental data.
The discrepancies demonstrate that there must be either some other unknown
production mechanisms or simply that perturbative QCD is not valid in this
case.

\begin{figure}[t]
\centering
\leavevmode
\epsfysize=200pt
\epsfbox{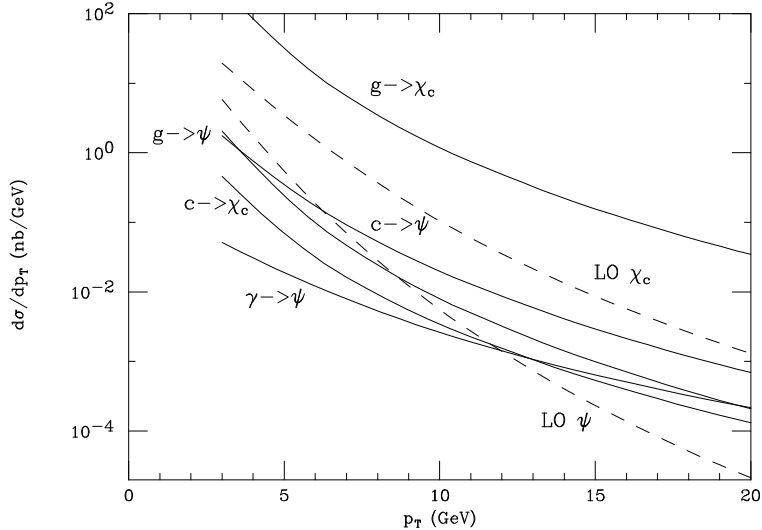}
\caption[]{\label{fig3} \small
Contributions to the differential cross section for inclusive $J/\psi$
production at the Tevatron: fragmentation into $\psi$ (solid curves), and
the leading order contributions (dashed curves). (Taken out from
Ref.~\cite{BDFM}.)
}
\end{figure}

\begin{figure}[t]
\centering
\leavevmode
\epsfysize=200pt
\epsfbox{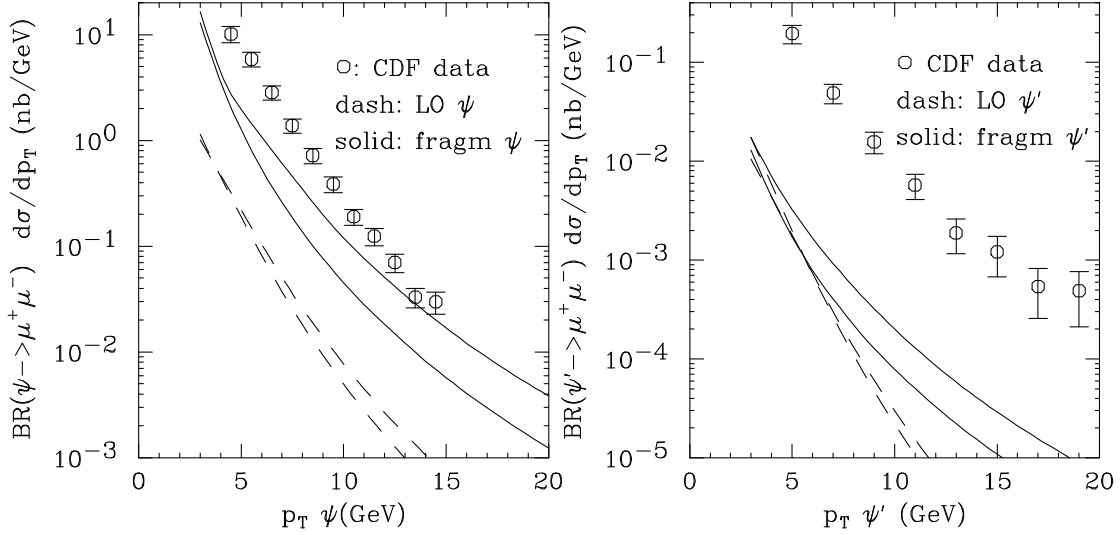}
\caption[]{
\label{fig4} \small
Preliminary CDF data for prompt $\psi$ and $\psi'$ production with theoretical
predictions of the total fragmentation contributions (solid) and the leading
order fusion contributions (dash).
}
\end{figure}

In order that perturbative QCD is still the means to understand the
production of heavy quarkonia, it is advantageous to consider higher order
contributions, which are more important than the lowest order fusion
mechanism.   It was shown explicitly in Ref.~\cite{BDFM} that
the contributions to $J/\psi$ and $\psi'$ production from gluon, charm quark,
and photon fragmentation are more important than the lowest order fusion
mechanisms beyond certain values of $p_T$ (see Fig.~\ref{fig3}.)
Among all the fragmentation contributions that are relevant to $J/\psi$
production, the largest one comes from gluon fragmentation into $\chi_{cJ}$
followed by the radiative decay $\chi_{cJ}\to J/\psi + \gamma$ \cite{gpfrag}.
The gluon fragmentation $D_{g\to \chi_{cJ}}(z)$ consists of two pieces,
one of which is the color-singlet part of order $\alpha_s^2$ and the other
piece is the color-octet part of order $\alpha_s$.
When the fragmentation contributions are included, the theoretical
prediction matches the experimental data within a factor of 2 -- 3
\cite{CG,BDFM,roy-1} (see Fig.~\ref{fig4}), which
is within the uncertainties from the mass of charm quark, the factorization
scale, higher order QCD corrections, and relativistic corrections.

While anomalous $J/\psi$ production seems to be solved, however, the data for
$\psi'$ production is still a factor of 20 -- 30 above the theoretical
prediction, even after including the fragmentation contributions (see
Fig.~\ref{fig4}). The $\chi_{cJ}(2P)$ states are predicted to be above the
$D\bar D$ threshold and therefore
do not contribute to $\psi'$ production.  This discrepancy is sometimes
referred as the $\psi'$ anomaly.  Of course, there have been speculative
solutions to the anomaly.  The most obvious solution is the hypothesis
that $\chi_{cJ}(2P)$ states are metastable such that they decay with
appreciable branching ratios into $\psi'$ \cite{cho,cho-spin,close,roy-2}.
According to potential models,
the $\chi_{cJ}(2P)$ states are above the $D\bar D$ threshold, but the
decay of $\chi_{cJ}\to D\bar D$ might be suppressed due to
a D-wave suppression.
Therefore, an appreciable fraction of $\chi_{cJ}(2P)$ can decay into
$\psi'$.  In order to explain the $\psi'$ anomaly, a branching  ratio
$B(\chi_{cJ}(2P) \to \psi' + \gamma) \approx 5-10$\% is needed.  However,
such a large branching ratio is unflavored by potential models.
There is also another mechanism due to Braaten and Fleming \cite{sean},
who proposed
that the production is via gluon fragmentation into a color-octet
$^3S_1^{(8)}$ state, which then nonperturbatively emits
a pair of gluons to make the transition into $\psi'$ state.  This
mechanism is suppressed by powers of $v$, which is the relative velocity
between $c$ and $\bar c$ inside the charmonium.
But on the other hand, this fragmentation
is flavored by two powers of $\alpha_s$ compared to the color-singlet
fragmentation function ($\alpha_s^3$).  It means that this fragmentation
mechansim could be potentially large because the corresponding
fragmentation function is only of order $\alpha_s$, though suppressed
by powers of $v$.  The major uncertainty is the determination of the
nonperturbative parameter associated with the soft emission of the
gluon-pair in color-octet  $0^+$ state.  This mechanism can be tested
rather easily because the $\psi'$ produced will be entirely transversely
polarized \cite{sean,cho-spin}, and the polarization can be easily
measured experimentally by looking at the angular distribution of the
muon pair in the decay of $\psi'$.  If this mechanism is tested to be
important, it will significantly affect all hadro-, photo-, and electro-
production of charmonium.
In fact, a very recent analysis indicated that only about one third of the
prompt $J/\psi$ comes from $\chi_{cJ}$ decays, while the rest is
{}from direct $J/\psi$ production.  It means that there is another
important production mechanism other than the gluon fragmentation into
$\chi_{cJ}$'s followed by the radiative decay of $\chi_{cJ}$'s into
$J/\psi$.   This might be the hint showing the importance of Braaten-Fleming's
color-octet mechanism \cite{sean} in $J/\psi$ production as well.

\section{A Fragmentation Model}

In this section I describe a fragmentation model \cite{spin,talks,pqcd} for
the fragmentation of heavy quarks into heavy-light mesons.  This model is
based on the PQCD fragmentation functions that are presented in
Eqs.~(\ref{dz1}) and (\ref{dz2}) for S-wave mesons, and in
Ref.~\cite{pbcfrag} for P-wave mesons.  But since most of the experimental
data are on S-wave states, I concentrate on Eqs.~(\ref{dz1})--(\ref{dz2}).
Let me rewrite the fragmentation functions here:
\begin{eqnarray}
\label{model}
D_{Q \rightarrow Q\bar q(^1S_0)}(z,\mu_0) & = &
N \, \frac{rz(1-z)^2}{(1-(1-r)z)^6}
\left[ 6 - 18(1-2r)z + (21 -74r+68r^2) z^2 \right. \nonumber \\
 && \left. -2(1-r)(6-19r+18r^2)z^3  + 3(1-r)^2(1-2r+2r^2)z^4 \right] \,,
\end{eqnarray}
\begin{eqnarray}
\label{model1}
D_{Q\rightarrow Q\bar q(^3S_1)}(z,\mu_0) & = &
3N\,  \frac{rz(1-z)^2}{(1-(1-r)z)^6}
\left[ 2 - 2(3-2r)z + 3(3 - 2r+ 4r^2) z^2  \right . \nonumber \\
&& \left.  -2(1-r)(4-r +2r^2)z^3  + (1-r)^2(3-2r+2r^2)z^4 \right] \,,
\end{eqnarray}
where $N$, instead of being given in terms of the wavefunction and the running
coupling constant, is now treated as a free parameter governing the
overall normalization together with another free parameter $r$,
$r$ being the mass ratio $m_q/(m_Q+m_q)$.
This model, in the limit $r\to 0$, approaches
 the fragmentation of a very heavy quark into a heavy-light meson.
In reality, the heavy quark is either $b$ or $c$, and the light component
is $u$, $d$, or $s$, and therefore $r$ is a small parameter.
In principle, when the lighter constituent quark becomes a light
quark, there is that nonperturbative physics involved in the
fragmentation process.  But we do expect our PQCD fragmentation functions
with $N$ and $r$ as free parameters can at least provide a qualitative
picture and hence a reasonable model for fragmentation
into heavy-light mesons.  This model is suitable for $b\to B,B^*, B^{**}$, ...
and $c\to D,D^*, D^{**}$, ... mesons.

This fragmentation model has certain advantages over previous models
\cite{peter} in
the literature.  First of all, this model lies on a firm basis of PQCD.
It is rigorously correct in the limit when $m_q \gg \Lambda_{\rm QCD}$
and higher order corrections can be systematically calculated.  The spirit
of our model is the continuation of $m_q$ to a value close to
$\Lambda_{\rm QCD}$.  The most obvious advantage of this model is the ability
to predict different results for different spin-orbital states with only
two parameters, in contrast to the Peterson model.
Another advantage is that the fragmentation functions for the same orbital
angular momentum share the same paramter $N$, as shown in Eqs~(\ref{model})
-- (\ref{model1}).   This is a substantial improvement when ratios of
the fragmentation functions are measured, in which the $N$ dependence cancels
out.
For example, our model can predict the ratio $P_V=V/(V+P)$ as a function of
$r$ only, where $V$ is the vector meson and $P$ is the pseudoscalar.

Another theoretical issue is that the PQCD fragmentation functions are
consistent with heavy quark symmetry, which I explain in more detail next.
According to an analysis using the heavy quark effective theory (HQET)
\cite{randall}, the fragmentation function for a heavy quark $Q$ into a
hadron $H_Q$ containing a single heavy quark $Q$ at the heavy quark mass
scale $m_Q$ is
\begin{equation}
D_{Q\to H_Q}(z) = \frac{1}{r} a(y) + b(y) + {\cal O}(r)
\end{equation}
which is a heavy quark mass expansion in $r=\frac{m_{H_Q} -m_Q}{m_{H_Q}}$
and $y$ is a rescaled variable of $z$, $y=\frac{1-(1-r)z}{rz}$.
The leading term is of order $1/r$, i.e., $m_Q$, while the next-to-leading
term is of order $r^0$.  The PQCD fragmentation functions in
Eqs.~(\ref{model}) and (\ref{model1}) can be expanded in powers of $r$ and
reexpressed in terms of $y$, as
\begin{equation}
\label{expansion}
\begin{array}{rcl}
D_{Q\to Q\bar q (^1S_0)}(z) &=& \frac{N(y-1)^2}{y^6}\left( \frac{1}{r}
(3y^2 + 4y +8) + (3y^3 +15y^2 +8y -8) +... \right) \\
D_{Q\to Q\bar q (^3S_1)}(z) &=& \frac{N(y-1)^2}{y^6} \left (\frac{3}{r}
(3y^2 +4y +8) -3(y^3 +y^2-8y+8) +  ... \right )
\end{array}
\end{equation}
The above expansions in powers of $r$ are consistent with the HQET analysis
\cite{randall} that the leading term is order $1/r$.  Actually, if examined
more carefully, the leading terms of the $^1S_0$ and $^3S_1$ expressions
are exactly in the ratio of 1:3, which is the value
 predicted by heavy quark spin
symmetry.  The next-to-leading terms are not in the ratio of 1:3, and
they explicitly break the spin symmetry.
This fact also prompted us to derive the PQCD fragmentation functions
independently from the HQET Lagrangian \cite{pqcd}. By using the leading
and the next-to-leading terms of the HQET Lagrangian, we succeeded in
obtaining the same results as the heavy quark mass expansions in
Eqs.~(\ref{expansion}).
Therefore, with the consistency with heavy quark symmetry and HQET we
have more confidence in applying our PQCD fragmentation functions as
a fragmentation model for heavy quark fragmenting into heavy-light
mesons, namely, $\bar b\to B,B^*, B^{**}$ and $c\to D,D^*,D^{**}$,...
When more data on P-wave mesons are available, comparisons with
the P-wave fragmentation functions can also be made.  For the moment the
data are more or less entirely on the S-wave states.  I shall demonstrate
a couple of  comparisons between the predictions by this model and the
experimental data.  We shall look at $P_V$ and $\langle z \rangle$.

\begin{figure}[t]
\centering
\leavevmode
\epsfysize=200pt
\epsfbox{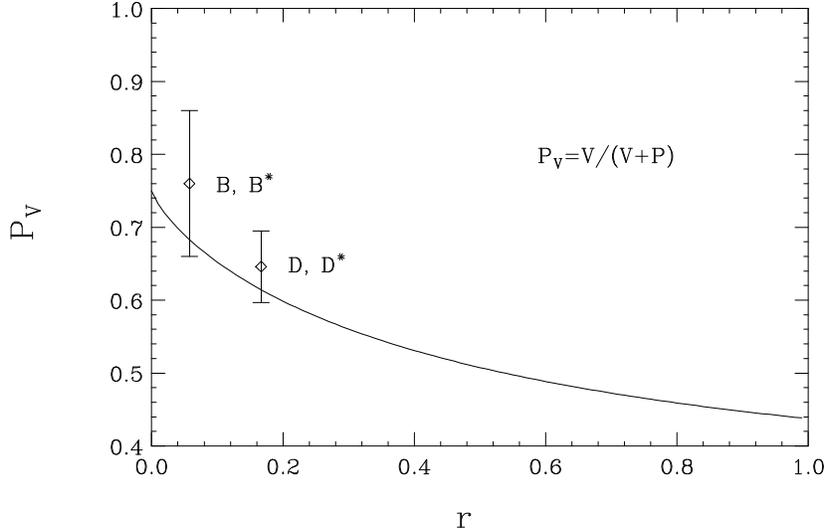}
\caption[]{
\label{fig5}\small
The ratio $P_V=V/(V+P)$, where $V$ denotes the vector state ($D^*,B^*$) and
$P$ denotes the pseudoscalar ($D,B$).  The data for $D,D^*$ system and
for $B,B^*$ system are shown.
}
\end{figure}

\noindent
(A) $P_V$ for the charm system is defined as
$P_V=D^*/(D+D^*)$, which is a measure of the population of $D^*$ in a
sample of $D$ and $D^*$ mesons.  Since fragmentation is the dominant
production mechanism for charm mesons and $D,D^*$ mesons dominate,
$P_V$ can be expressed in terms of the fragmentation functions
\begin{equation}
P_V = \frac{ \int dz D_{c\to D^*}(z)}{\int dz D_{c\to D}(z) +
                                      \int dz D_{c\to D^*}(z) }\;,
\end{equation}
which is a function of $r$ only.  The prediction by the PQCD fragmentation
model is shown by the solid curve in Fig.~\ref{fig5}.  At $r=0$, the
heavy quark mass limit, $P_V=0.75$, which is exactly the value given
by the naive spin counting.  At $r>0$ $P_V$ is always smaller than
0.75, which implies that $D^*$ is produced less than given by heavy quark
spin symmetry.  This can be understood in terms of the mass splitting
between $D$ and $D^*$ mesons, which can be accounted for by the
$\sigma^{\mu\nu}G_{\mu\nu}/M$ term in the HQET.

A compilation of data on $P_V$ can be found in Ref.~\cite{peskin},
in which the updated branching ratio $B(D^{*+} \to D^0 \pi^+)=0.681\pm
0.016$ was used, and the average $P_V=0.646 \pm 0.049$.  This value
of $P_V$ also indicates that $D^*$ mesons are produced less than it
should be as given by heavy quark spin symmetry.  For the charm quark
we take $m_c=1.5$ GeV and the light constituent quark mass to be 0.3
GeV, therefore $r=\frac{m_{\rm light}}{m_c + m_{\rm light}}=0.167$.
The data is then plotted on the graph and very good agreement
is obtained.  Recently, the data for $B,B^*$ system has also been
available with $P_V(B,B^*)= 0.76 \pm 0.08 \pm 0.06$ \cite{L3}.
We take $m_b=4.9$ GeV and
$m_{\rm light}=0.3$ GeV again, therefore $r=0.06$.  The prediction
is still less than 1$\sigma$ from the data point.  From the
figure we can see that if we choose a smaller value of $m_{\rm light}$,
say 0.2 GeV, we could even get a better agreement.  The errors in $P_V$
certainly allow us to vary $m_{\rm light}$ more than 0.15 GeV such that
the prediction is still within $1\sigma$.  Or, we can use the
experimental value of $P_V$ to fix the parameter $r$.

\begin{figure}[t]
\centering
\leavevmode
\epsfysize=200pt
\epsfbox{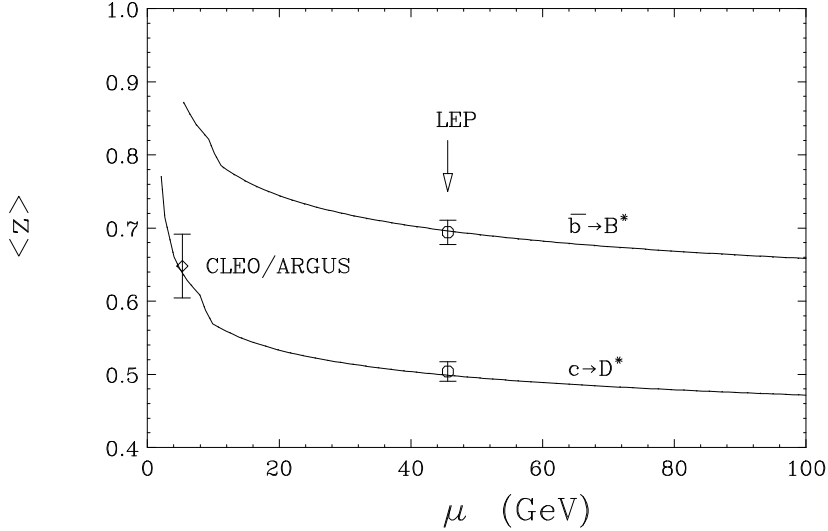}
\caption[]{
\label{fig6} \small
The average $\langle z \rangle^\mu$ for $c\to D^*$ and for $\bar b\to B^*$
fragmentation versus the scale $\mu$.  The experimental measurements from
LEP ($\mu=m_Z/2$) and from CLEO/ARGUS ($\mu=5.3$ GeV) are shown.
}
\end{figure}

\vglue 0.4cm
\noindent
B) $\langle z \rangle$ is the average longitudinal momentum fraction that is
transferred from the heavy quark to the meson.  In terms of fragmentation
functions, $\langle z \rangle^{\mu}$ at a scale $\mu$ is given by
\begin{equation}
\label{z}
\langle z \rangle^{\mu}_{c\to D^*} = \frac{\int dz\, zD_{c\to D^*}(z,\mu)}
{\int dz \, D_{c\to D^*}(z,\mu)}\;.
\end{equation}
Experimentally,  the inclusive $c\to D^*$ channel was measured at LEP,
at CLEO,  and at ARGUS.
The $\langle z \rangle^{\mu}_{c\to D^*}$ given in Eqn.~(\ref{z}) is the ratio
of the second to the first moments of the fragmentation function at the scale
$\mu$.  Since the anomalous dimensions of the moments are known explicitly,
the scaling behavior of $\langle z \rangle^{\mu}$ can be determined to be
\begin{equation}
\label{evol}
\langle z \rangle^{\mu} = \langle z \rangle^{\mu_0} \left(
\frac{\alpha_s(\mu)}{\alpha_s(\mu_0)} \right )^{\frac{-2\gamma}{b}} \;,
\end{equation}
where $\gamma=-4C_F/3$, $C_F=4/3$, $b=(11N_c - 2 n_f)/3$, $N_c=3$, $n_f$ is
the number of active flavors at the scale $\mu$, and $\langle z
\rangle^{\mu_0}$ is the value determined at the initial scale $\mu_0$.
Taking the inputs: $m_c=1.5$ GeV, $m_{u,d}=0.3$ GeV, $\mu_0=m_c+2m_{u,d}=2.1$
GeV, we have $r=0.167$, and $\langle z \rangle^{\mu_0}_{c\to D^*}=0.77$ and
$\langle z \rangle^{\mu=m_Z/2}_{c\to D^*}=0.50$.  The variation of
$\langle z \rangle^{\mu}_{c\to D^*}$ and $\langle z \rangle^{\mu}_{b\to B^*}$
as functions of $\mu$ are shown in Fig.~\ref{fig6}, where we chose $m_b=4.9$
GeV.   The curve for $\langle z \rangle_{b\to B^*}^\mu$ was also shown
because we want to demonstrate that
using the same $m_{u,d}$ and $m_b=4.9$ GeV,  the results predicted also agree
with the data for the bottom quark fragmentation at LEP.

The measured quantity is $\langle x_E \rangle$, which is the energy of
the meson relative to one half of the center-of-mass energy of the machine,
and $x_E$ should be a good approximation to $z$.
At LEP, the average value of $\langle x_E \rangle_{c\to D^*} = 0.504
\pm 0.0133$ \cite{lep-c}.
For the bottom quark, only the inclusive hadron production has been measured.
But we expect that $\langle x_E \rangle_{b\to B^*}$ should be close to
$\langle x_E \rangle_{b\to H_b}$, where $H_b$ is a bottom hadron, because the
$b\to B^*$ is the dominant fragmentation mode of the bottom quark.
The average value of $\langle x_E \rangle_{b\to H_b}=0.694\pm 0.0166$
\cite{lep-b}.
Also, we have data on $c\to D^*$ from CLEO and ARGUS \cite{cleo2}.
Combining the CLEO and ARGUS  data we have
$\langle x_E \rangle_{c\to D^*}=0.648\pm0.043$.  The scale of the measurements
is taken to be one half of the center-of-mass energy of the machines, so
it is $m_Z/2$ at LEP and 5.3 GeV at CLEO/ARGUS.  These data are shown in
Fig.~\ref{fig6}.  Excellent agreement is demonstrated.
The only inputs to these comparisons are simply $r$ and $\mu_0$.
Once they are fixed, $\langle z \rangle^{\mu_0}$ can be calculated by
Eqn.~(\ref{z}) and evolved by Eqn.~(\ref{evol}) to any scale $\mu$.
The results show agreement at two different scales.  This is a big
contrast to Peterson fragmentation model, which fits to different values of
mass ratio $\epsilon_Q$ at different scales.

\section{Conclusions}

In this proceedings I have summarized some recent work on
PQCD fragmentation functions.
The $\psi$ production at the Tevatron is the first evidence showing the
importance of parton fragmentation in quarkonium production in the large
$p_T$ region.  I have also shown the results of the production
of $\bar bc$ mesons by the fragmentation approach.
Finally, a fragmentation model based on the PQCD fragmentation functions
is advocated to describe the fragmentation of heavy quarks into
heavy-light mesons.  This model lies on a firm basis of PQCD and is
consistent with HQET, and was successfully applied to fit data on $P_V$
and $\langle z \rangle^\mu$.
Other work includes the estimation of the strange-quark mass parameter
$m_s$ in the $B_s$ and $B_s^*$ system, in which the probability of a $b$
quark going into a stranged $B$ meson is fitted to the value of $m_s$
\cite{bs}, and a value of  about 300 MeV was obtained.
 There was also a calculation \cite{tc-heli} of the Falk-Peskin variable
$w_{3/2}$ \cite{peskin} using the PQCD fragmentation functions.
Another work was the  extension to the fragmentation
of a heavy quark into a baryon containing two heavy quarks \cite{baryon}.

\section*{Acknowledgement}
I am indebted to my collaborators, Eric Braaten, Tzu Chiang Yuan, and
Sean Fleming on the subjects presented in this proceedings.  I also thank
Adam Falk for inviting me to give the talk.
This work was supported by the DOE-FG03-93ER40757.


\end{document}